\documentclass[prd,nofootinbib,
amsmath,amssymb,aps,twocolumn,
]{revtex4-1}

\usepackage[colorlinks=true
,urlcolor=blue
,anchorcolor=blue
,citecolor=blue
,filecolor=blue
,linkcolor=blue
,menucolor=blue
,linktocpage=true
,pdfproducer=medialab
,pdfa=true
]{hyperref}
\usepackage{hyperref}
\usepackage{graphicx}
\usepackage{bm}

\begin{document}
\title{Thermal Sunyaev-Zel'dovich anisotropy due to primordial black holes}
\author{Katsuya T. Abe}
\email{abe.katsuya@e.mbox.nagoya-u.ac.jp}
\author{Hiroyuki Tashiro}
\author{Toshiyuki Tanaka}
\affiliation{Division of Particle and Astrophysical Science, Graduate School of Science, Nagoya University, Chikusa, Nagoya 464-8602, Japan}


\begin{abstract} 
We investigate the thermal Sunyaev-Zel'dovich~(SZ) effect caused by primordial
black holes~(PBHs) on the cosmic microwave background~(CMB) temperature fluctuations.
The gas accreting on a PBH heats up by the release of the gravitational energy. 
As a result, the heated gas in the vicinity of the PBH emits UV and x-ray photons. These photons can ionize and heat the intergalactic medium~(IGM) around the PBH. 
Assuming the simple model of these emitting photons,
we compute the profiles of the IGM ionization fraction and temperature around a PBH by using the numerical calculation of the radiative transfer.
Using these profiles, we evaluate the Compton $y$-parameter created by
the IGM gas around a PBH.
Finally, we estimate the CMB temperature angular power spectrum due to the
PBH SZ effect in our model.
We show that the SZ temperature anisotropy due to the PBHs has the flat
angular power spectrum on small scale, $l\leq2000$ and
could dominate the primordial temperature spectrum on smaller scales
than the Silk scale.
This flat spectrum extends to the scale of the ionized region by the PBH emission.
Therefore, in future SZ anisotropy measurements, the detection or nondetection of the flat spectrum gives useful information about the existence of PBHs.
\end{abstract}
\maketitle

\section{Introduction}
Many astronomical observations
including the motion of stars in a galaxy, the large-scale structure
distribution, and the cosmic microwave background~(CMB) temperature anisotropy
indicate the existence of dark matter~(DM) in the Universe.
However, its nature is still poorly understood.
Currently, since the presence of DM can be probed only through
its gravitational interaction,
observational information about its nature is very limited.
However, there are many works to study the nature of
DM and many theoretical models have been suggested so far.
Generally, the DM candidates are classified into two types;
the nonbaryonic particle model and astrophysical compact object model.
A lot of models for the nonbaryonic particle type
are predicted in the physics beyond the standard model such as
weakly interacting massive particles~\cite{WIMP}, axions~\cite{axion1,axion2,axion3,axion4} and axionlike particles~\cite{ALP1}.
On the other hand, in the latter type,
a primordial black hole~(PBH) is the most potential candidate~\cite{Carr&Kuhnel_pbhasdm}.

PBHs could have formed from high-density peaks in the very early
Universe~\cite{Zeldovich&Novikov_whatispbh,Hawking_pbhform}.
In the inflation paradigm, the density fluctuations are generated from
the quantum fluctuations. When
an overdense region which exceeds a critical density threshold
enters the horizon scale, the gravitational collapse of this region happens to form a PBH.
The resultant mass of the formed PBH corresponds the horizon mass at the horizon-crossing epoch of the overdense region. As a result, the
PBH mass range can span very widely.
Therefore,
the PBH abundance has been studied for a long time
not only for DM but also as the relic of the primordial density fluctuations on small scales.
Besides, the recent detection of gravitational wave~(GW) events
drew attention to PBHs as sources of GWs.
The analysis of the observed GW data reveals that
the detected GW events were
produced by binary black hole mergers with masses larger than $20~M_{\odot}$~\cite{LIGO_2016GW150914,LIGO_2017GW170104,LIGO_2017GW170814}.
It would be difficult to produce a black hole with such large mass
from stellar evolution
in the standard solar metallicity environment~\cite{2016Natur.534..512B}.
On the other hand, the broad mass range of PBHs can cover the observed black hole
masses.
Therefore, the PBH abundance is also studied as a responsible source for
the detected GW events~\cite{Bird_gwpbh,Sasaki&Suyama_pbhgw2016,2017PDU....15..142C}.

The existence of PBHs affects various cosmological phenomena, depending
on the PBH mass.
For small mass PBHs,
the abundance can be constrained by the effects of their evaporation.
As first pointed out by Hawking~\cite{Hawking_radiation},
a black hole emits many kinds of particles with the thermal spectrum.
As a result, PBHs with a mass smaller than $10^{15}~$g have evaporated by
the present epoch. The abundance constraint on evaporated PBHs is
obtained
by investigating the effect of their evaporation on 
big bang nucleosynthesis~\cite{Carr&Kohri_newconstraintonPBH}, the CMB spectrum distortion~\cite{Tashiro&Sugiyama_cmbpbh}, the recombination
and reionization processes~\cite{2017PhRvD..95h3006C,2018PhRvD..98d3006C}, and the
diffuse gamma-ray background~\cite{1976ApJ...206....1P,1976ApJ...206....8C,1991ApJ...371..447M}.

For nonevaporated PBHs, the robust constraint is provided by
gravitational lensing observations~\cite{2007A&A...469..387T,Niikura_gralensing_pbhconstraint,2018PhRvL.121n1101Z}.
The black hole merger rate evaluated from the recent
detection of GW events also constrains the PBH abundance~\cite{Sasaki&Suyama_pbhgw2016,2017PDU....15..142C}.
Recently several works focused on the gas
accreting on massive PBHs. Because of the release of the gravitational
energy during accretion, the gas becomes hot 
and emits x-ray and UV photons~\cite{ROM,Kamionkowski&Alihaimoud_pbhlimit,Poulin:2017bwe}. 
Resultantly the surrounding gas around a PBH 
is heated and ionized. 
Studying the cosmological effects of such heating 
and reionization provides the constraint on 
stellar mass PBHs with the recent CMB
measurement~\cite{ROM,Kamionkowski&Alihaimoud_pbhlimit,Poulin:2017bwe,2017JCAP...10..052B}.
There are also some works suggesting that
future 21-cm observations can probe these PBH heating
and ionizing processes and give the strong constraint
on PBHs in this mass range~\cite{Tashiro&Sugiyama_21cmpbh,Kitajima_21cm}.

In this paper, we study the thermal Sunyaev-Zel'dovich~(SZ) effect due to
PBHs. The thermal SZ effect is the distortion of the CMB energy spectrum
through the inverse Compton scattering by high-energy electrons in hot
plasma and this effect is parametrized by the Compton $y$-parameter~\cite{Zeldovich&Sunyaev_yandT,Sunyaev&Zeldovich_SZ2}.
As mentioned above, PBHs can heat and ionize the surrounding gas.
The resultant ionized gas contributes not only to the global optical depth
of the Thomson scattering as discussed in Refs.~\cite{ROM,Kamionkowski&Alihaimoud_pbhlimit,Poulin:2017bwe},
but also to creating the nonzero Compton $y$-parameter. The sky-averaged~(global) Compton $y$-paremter is already evaluated in Ref.~\cite{Kamionkowski&Alihaimoud_pbhlimit}. 
They have obtained the global Compton $y$-parameter induced by PBHs
from evaluating the energy deposit of the PBH luminosity into a plasma.
According to their estimation, the global Compton $y$-parameter
could be roughly $y \sim 10^{-7} f_{\rm PBH}$, where $f_{\rm PBH}$ describes
the PBH abundance fraction to the total DM one. Since this estimated value is below the sensitivity of the Far-Infrared Absolute Spectrophotometer~\cite{FIRAS},
they concluded that 
it is difficult to obtain
a constraint on the PBH abundance
from the global Compton $y$-parameter 
with the current observation status.

Here, we focus on the spatial fluctuations of the Compton $y$-parameter induced by PBHs
and the resultant CMB temperature anisotropy due to them.
The angular power spectrum of the CMB temperature anisotropy
could provide additional information through its amplitude and shape depending on scales
which we cannot access from
the global value.
Our main aim is to study the dependence of the SZ angular power spectrum on the PBH properties such as the abundance, mass and so on.

To evaluate the spatial fluctuations of the Compton $y$-parameter, it is required 
to obtain the profiles of the ionization faction and the intergalactic medium~(IGM) gas temperature around a PBH. Introducing the emission efficiency parameter of the PBH luminosity,
we evaluate these profiles
through the one-dimensional radiative transfer calculation.
Then,
assuming that the PBH distribution follows the dark matter density
fluctuations,
we calculate the CMB temperature anisotropy due to the SZ effect of PBHs.
We investigate the dependence of the PBH fraction, the PBH mass and 
the emission efficiency parameter on the generated CMB anisotropy.
From the comparison with the small-scale measurement of the CMB
temperature anisotropy by South Pole Telescope~(SPT)~\cite{SPT},
we also discuss the suggestion of the SZ effect due to PBHs on the constraint on the PBH abundance with
the emission efficiency parameter.

The rest of this paper is organized as follows. 
In Sec.~II, we compute the gas temperature
and ionization fraction around a PBH.
Accordingly,
we calculate the profile of the thermal SZ effect.
In Sec. III,~introducing the PBH fraction to dark matter,
we evaluate the CMB temperature angular power spectrum
due to the thermal SZ effect around PBHs.
We also discuss the PBH abundance constraint 
from comparing our results with the SPT data.
Finally, 
we summarize in Sec.~IV.
Throughout our paper,
we take the flat $\Lambda$CDM model with the Planck best fit parameters~\cite{Planck2018_cospara}:
$(\Omega_{\rm m},\Omega_{\rm b},h,n_{\rm{s}},\sigma_{8})$=$(0.32,0.049,0.67,0.97,0.81)$. 

\section{THERMAL SUNYAEV-ZEL'DOVICH EFFECT DUE TO A PBH}
\label{section2}

In this section, we evaluate the gas temperature and ionization fraction
of the IGM around a PBH,
assuming the photon energy spectrum emitted from the hot gas in the vicinity of a PBH.
Accordingly, we calculate the profile of the thermal SZ effect.
\subsection{The luminosity from a PBH}

Since a PBH creates a gravitational potential, 
the surrounding gas accretes on the PBH. 
During accretion, as it goes closer to a PBH, 
the gas becomes hot enough to emit x-ray and UV
photons~\cite{ROM,Kamionkowski&Alihaimoud_pbhlimit,Poulin:2017bwe}.
However, since the gas temperature highly depends on the astrophysical
processes and environmental condition, there is a theoretical uncertainty in this luminosity.
References~\cite{ROM,Kamionkowski&Alihaimoud_pbhlimit} evaluated the luminosity of a PBH considering 
spherical accretion on the PBH
and showed that
the luminosity is much lower than the Eddington luminosity.
On the other hand,
Ref.~\cite{Poulin:2017bwe} pointed out that a PBH could have an accretion disk and that its luminosity is much higher
than in the case of the spherical accretion.
Moreover, when a massive PBH has an accretion disk, its luminosity might become nearly sub-Eddington luminosity similar to the active galactic nuclei as considered in Ref.~\cite{2018JCAP...05..017B}.

Therefore, this emission efficiency can depend on the redshift, the PBH
mass and other physical conditions
around the PBH. 
Here, for simplicity, we introduce
one free parameter for the emission efficiency, $\epsilon$, which represents
the total PBH luminosity, $L_{\rm PBH}$,
in terms of the Eddington luminosity.
\begin{equation}
L_{\rm{PBH}} =
\epsilon L_{\rm{Edd}},
\end{equation}
where $L_{\rm{Edd}}$ is the Eddington luminosity for mass $M$
\begin{equation}
L_{\rm{Edd}}=3.2 \times 10^4 L_\odot (M/M_\odot).
\end{equation}

We also assume the power-law type of luminosity spectrum:
\begin{equation}
L_{\rm{PBH},\nu} ={\cal A} \nu^{-1.5},
\label{eq:fre_spe}
\end{equation}
where $\cal A$ is determined by
\begin{equation}
L_{\rm{PBH}} = \int_{\nu_{\rm L}} L_{{\rm PBH},\nu} d\nu,
\end{equation}
with the Lyman limit frequency, $\nu_{\rm L}$.
In Eq.~\eqref{eq:fre_spe}, we take the frequency spectral index of $-1.5$
similar to that in the case of galactic black holes~\cite{Giannious_blackholephotonindex}.

\subsection{IGM temperature and ionization profiles around a PBH}

Now we consider the profiles of the ionization fraction and gas temperature 
around a PBH with $L_{\rm PBH}$.
For simplicity, we consider only hydrogen as the IGM gas component.

The evolutions of the ionization fraction,
$x_{\rm{e}}$ and 
temperature $T_{\rm{gas}}$ in the IGM 
are given by the following equations,
\begin{align}
&\frac{dx_{\rm{e}}}{dt}=k_{\rm{HI, \gamma}}-\alpha_{\rm{B}}n_{\rm{H}}x_{\rm{e}}^{2},
\label{xh1}\\
&\frac{d\rm{T}_{\rm{gas}}}{dt}=(\gamma-1)\frac{\mu m_{\rm{p}}}{k_{\rm{B}}\rho}\left(\frac{k_{\rm{B}}T_{\rm{gas}}}{\mu m_{\rm{p}}}~\frac{d\rho}{dt}+\Gamma-\Lambda\right),
\label{temp}
\end{align}
where $n_{\rm{H}}$, $k_{\rm{B}}$,  $\gamma$, $\mu$, $m_{\rm{p}}$, $\rho$
and $\alpha_{\rm{B}}$ are the number density of hydrogen nucleus, the
Boltzmann constant, the IGM gas adiabatic index, $\gamma =5/3$,
the mean molecular weight, the proton mass, the gas mass density,
and the case B recombination rate given in Ref.~\cite{Fukugita&Kawasaki_cooling} respectively.  
In Eqs.~\eqref{xh1} and \eqref{temp}, 
$k_{\rm{HI, \gamma}}$, $\Gamma$ and $\Lambda$ are the ionization, heating
and cooling rates, respectively.
Here, we adopt cooling mechanisms due to recombination cooling, collisional ionization cooling, collisional excitation cooling and Compton cooling~\cite{Fukugita&Kawasaki_cooling}.
For simplicity, we assume that the IGM gas is homogeneous and
we set the IGM gas density to the
cosmological mean~(background) values at a given time~$t$.

We consider only photons emitted from a PBH as the ionization and
heating source.
We solve Eqs.~\eqref{xh1} and \eqref{temp}
with the spherically symmetric assumption.
Since we evaluate the ionization and temperature evolutions for the cosmological
timescale, it is useful to introduce the comoving radial
distance from a PBH for representing the spatial position, instead of
the physical distance.
The ionization and heating rates at the comoving radial distance $r$
can be written, respectively, as
\begin{align}
k_{\rm{HI,
 \gamma}}(r)&={(1-x_{\rm{e}}(r))}\int^{\infty}_{\nu_{\rm{L}}}
{\cal F}_\nu(r)
\frac{\sigma_{\rm{HI,\nu}} }{h\nu}
 ~d\nu,
\label{kh1} \\
\Gamma(r) &={n_{\rm{HI}}(r)} \int^{\infty}_{\nu_{\rm{L}}}
{\cal F}_\nu(r)
\frac{(\nu-\nu_{\rm{L}})}
 {\nu}
 \sigma_{\rm{HI,\nu}}~d\nu,
\end{align}
where $h$, $n_{\rm HI}(r)$ and
$\sigma_{\rm{HI,\nu}}$ 
are the Planck constant, 
the neutral hydrogen number density given in $n_{\rm HI}(r) = (1-x_{\rm e}(r))n_{\rm H}$,
and the absorbed cross section area of ionization
photons,
$\sigma_{\rm{HI,\nu}}=6.3\times10^{-18}(\nu/\nu_{\rm{L}})^{-3}{\rm  cm}^{2}$, respectively.
In the above equations,
${\cal F}_\nu(r)$ represents the photon energy flux for a frequency~$\nu$
at a comoving distance~$r$:
\begin{equation}
{\cal F}_\nu(r)=
  \frac{L_{\rm{PBH,\nu}}}{4\pi
  a^{2}(t)r^{2}}
e^{-\tau_{\rm{HI,\nu}}(r)}.
\end{equation}
Here $a(t)$ is the scale factor normalized as $a(t_0) =1$ at the present
epoch, $t_0$, and 
$\tau_{\rm{HI,\nu}}(r)$ is the optical depth of HI gas from the central
PBH to the comoving distance~$r$,
\begin{equation}
\tau_{\rm{HI,\nu}}(r)=\int^{r}_{0} a(t) n_{\rm{HI}}(r')\sigma_{\rm{HI,\nu}}~dr'.
\end{equation}

Solving Eqs.~\eqref{xh1} and \eqref{temp} consistently
by a spherically symmetric radiative transfer code,
we obtain the evolutions of $x_{\rm e}$ and $T_{\rm gas}$ at each comoving
distance~$r$ and cosmic time~$t$.

\begin{figure}[tbp]
\centering
\includegraphics[width=8cm]{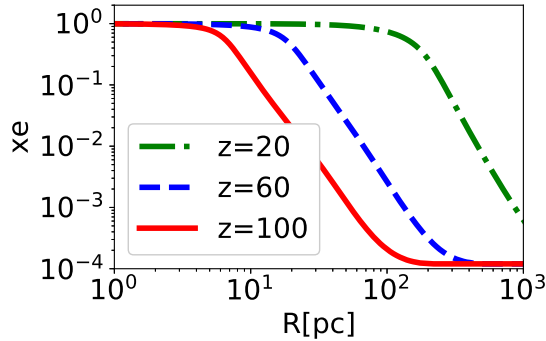}
\caption{The radial ionization fraction profile around a PBH
as a function of the physical distance from the PBH.
We set $M=10 M_{\odot}$ and $\epsilon=0.0001$ in this figure.
The red solid, blue dashed and green dashed-dotted lines
represent the profiles at $z=100$, $z=60$ and $z=20$, respectively. 
}
\label{xe} 
\end{figure}

Figure.~\ref{xe} shows the radial profile of the ionization fraction
as a function of the physical distance,~$R=a(t)r$, from a PBH.
In the figure,
we set $M=10~{M}_{\odot} and \epsilon=0.0001$.
The difference of colors represents different redshifts.
As the Universe evolves, the ionized region extends outward.
This
expansion is proportional to $(1+z)^{-2}$. This behavior is consistent with
the redshift evolution of the Str\"omgren radius with the constant
photon flux from a PBH.
The~Str\"omgren radius~$R_{\rm{s}}$~(in physical)
can be related to 
the number rate of photons emitted from a PBH,~$N_{\rm{PBH,\gamma}}$,
and has the redshift dependence as in
\begin{equation}
 R_{\rm{s}}=\left(\frac{3N_{\rm{PBH,\gamma}}}{4\pi n_{\rm e}n_{\rm p}\alpha_{\rm B}}\right)^{1/3}\propto (1+z)^{-2},\label{stromgrem_radius}
\end{equation}
where $N_{\rm{PBH,\gamma}}$ is obtained
from
\begin{equation}
N_{\rm{PBH,\gamma}}
=\int^{\infty}_{\nu_{\rm{L}}}d\nu \frac{L_{\rm{PBH,\nu}}}{h\nu}.
\label{npbh_Lpbh}
\end{equation}
Since we assume that the photon emission from a PBH does not have the
redshift evolution,
the redshift dependence in Eq.~\eqref{stromgrem_radius}
comes from the (physical) number density of electrons and protons,~$n_{\rm e},n_{\rm p}\propto (1+z)^{3}$.

\begin{figure}[tbp]
\centering
\includegraphics[width=8cm,clip]{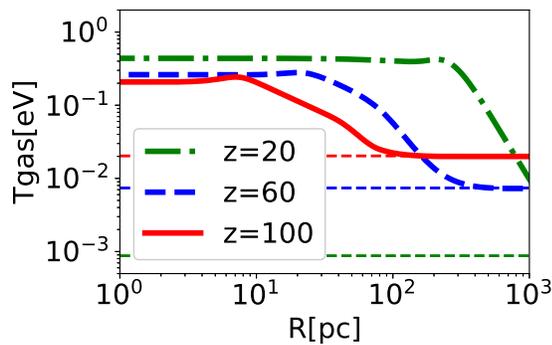}
\caption{
The radial temperature profile around a PBH
as a function of physical distance from the PBH.
We set $M=10 M_{\odot}$ and $\epsilon=0.0001$ in this figure.
The meaning of the color is the same as in Fig.~\ref{xe}.
The thin dashed lines represent the background IGM gas temperature
at each redshift.
}
\label{tgas} 
\end{figure}

Figure~\ref{tgas} provides the redshift evolution
of the IGM gas temperature.
We plot the radial profile of the temperature
as a function of the physical radial distance from a PBH.
The color difference shows the difference of the redshifts.
Similarly to the ionization fraction in~FIG.~\ref{xe},
the lower the redshift becomes, 
the more the volume of the heated region increases.
In the fully ionized central region,
the temperature profile is flat
and the redshift dependence of the amplitude is very weak.
Since the heating mechanism is photoionization,
the heating becomes efficient in the neutral region.
Therefore, the peak of the gas temperature locates at the spatial
position where the ionization fraction becomes 
about $x_{\rm e}\simeq 0.1$.
Thus, the heated region extends more than the ionized region.
In a sufficiently distant region, the gas temperature becomes the background IGM temperature
which is represented by the thin dashed lines in~Fig.~\ref{tgas}.

At the end of this subsection, we comment on the 
dependence of the ionization fraction and gas temperature on
the PBH luminosity, $L_{\rm PBH} \propto \epsilon M$.
According to Eq.~\eqref{stromgrem_radius} with~Eq.~\eqref{npbh_Lpbh},
the Str\"omgren radius depends on~$(\epsilon M)^{1/3}$.
As shown in Figs.~\ref{xe} and~\ref{tgas},
the profiles of $x_{\rm e}$ and $T_{\rm gas}$ 
is sensitive to
the evolution of the Str\"omgren radius.
Therefore, as $\epsilon$ or $M$ increases,
the ionization front expands towards outside
roughly proportional to $(\epsilon M)^{1/3}$.
However, since the temperature is determined by the balance between
the heating and cooling,
the amplitude of the temperature at the fully ionized region
does not have the simple
dependence on $\epsilon M$.


\subsection{Compton $y$-parameter induced by a PBH}
\label{sz_by_apbh}

Figures.~\ref{xe} and~\ref{tgas} 
show that a PBH can make a hot gas plasma around itself.
Passing through this plasma,
CMB photons suffer from the SZ effect.
As a result, the observed brightness temperature
of the CMB is shifted by $\Delta T_\nu$ 
at a frequency~$\nu$
from the background CMB temperature, $T_{\rm
CMB}$~\cite{Zeldovich&Sunyaev_yandT}.
At the comoving distance~$b$ from the PBH on the sky,
this observed shift of the brightness temperature
can be written with 
the Compton $y$-parameter,
\begin{equation}
\frac{\Delta T_\nu (b)}{T_{\rm CMB}}=
g(\nu) y (b), 
\end{equation}
where $g(\nu)$ is the frequency spectral function of the SZ effect,
\begin{equation}
g(\nu) = \frac{h\nu}{k_{\rm{B}}T_{\rm{gas}}}
 {\tanh}^{-1}\left(
 \frac{h\nu}{2k_{\rm{B}}T_{\rm{gas}}}\right).  
\end{equation} 
The Compton $y$-parameter at the comoving distance~$b$ on the sky,~$y(b)$, can be obtained
from the integral form along a line-of-sight with the impact
parameter~$b$
from the PBH,
\begin{equation}
y(b)= \int dx
\frac{\sigma_{\rm{T}} n_{\rm{H}}x_{\rm{e}}(\ell)}{m_{\rm{e}}c} 
 k_{\rm{B}}T_{\rm{gas}}(\ell), 
\label{yparam}
\end{equation}
where 
$\sigma_{\rm{T}}$ is the Thomson cross section, $m_{\rm{e}}$ is
the electron mass and $c$ is the speed of light.
In Eq.~\eqref{yparam},
$x$ is the comoving distance projected on the line-of-sight direction and 
$\ell$ is the comoving radial distance from the PBH satisfying, $\ell^2 = b^2 +x^2$.

\begin{figure}[tbp]
\centering
\includegraphics[width=8cm,clip]{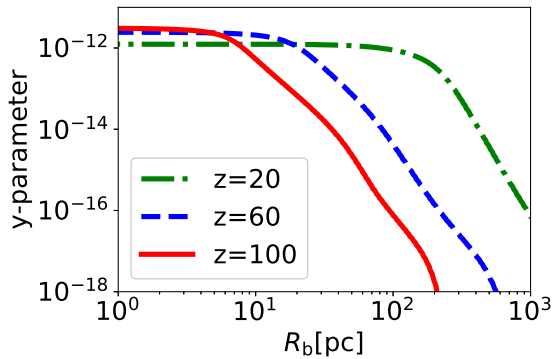}
\caption{
The Compton $y$-parameter as a function of the physical impact parameter 
from the PBH. 
We set $M=10 M_{\odot}$ and $\epsilon=0.0001$ in this figure.
The meaning of the color is the same as in Fig.~\ref{xe}.
}
\label{y} 
\end{figure}

Using Eq.~\eqref{yparam} with the profiles of the ionization
fraction and temperature shown in Figs.~\ref{xe} and~\ref{tgas},
we can compute the $y$-parameter.
We plot the $y$-parameter due to a PBH
with $M=10 M_{\odot}~and~\epsilon=0.0001$
as a function of the physical distance,~$R_{\rm b}=a(t)b$,
for different redshifts
in Fig.~\ref{y}.
As expected, the $y$-parameter profile depends on the ionization
fraction profile.
Therefore, when $R_{\rm b}$ becomes larger than the Str\"omgren radius,
the $y$-parameter quickly falls down. 
Following the evolution of the Str\"omgren radius, as the Universe evolves, the flat region in the $y$-parameter profile
extends outward.
In Eq.~\eqref{yparam},
most of the contribution to the integration
come from 
the region inside the Str\"omgren radius~(the fully ionized region).

The amplitude of $y$-parameter decreases as the Universe evolves
because the redshift dependence of the electron number density dominates that of the integral distance and gas temperature.

We also comment on the dependence of the $y$-parameter profile on~$\epsilon$ and~$M$.
As mentioned above, the $y$-parameter profile strongly depends on the Str\"omgren
radius. Since the Str\"omgren radius is proportional to $(\epsilon M)^{1/3}$,
the parameters $\epsilon$ and $M$ can affect the $y$-parameter
profile.
As $\epsilon$ or $M$ becomes large, the amplitude of the $y$-parameter
increases and the tail of the profile moves toward large $R_{\rm b}$.

Based on Fig.~\ref{y},
we can roughly estimate the total $y$-parameter induced by PBHs on the line-of-sight direction
between $z=100$ and $z=20$.
Figure~\ref{y} shows that
one PBH with~$M=10 M_{\odot}$
and~$\epsilon=10^{-4}$
provides $y \simeq 10^{-12}$.
Since the comoving number density of such PBHs
is about $n_{\rm PBH}
\sim 10^9 f_{\rm PBH}~\rm{Mpc}^{-3},$
where $f_{\rm PBH}$
is the PBH fraction 
to the total DM abundance,
the resultant typical distance of PBHs is roughly $10^{-3}\rm{Mpc}$.
Since the radial distance between $z=100$ and $z=20$
is about $10^3~\rm{Mpc}$,
the PBH number on the line-of-sight is 
$N_{\rm{PBH}}\sim 10^{6}.$
Therefore, the total $y$-parameter
integrated over the line-of-sight is 
$y \sim 10^{-6}$.
Consequently, this y-parameter can generate a SZ signature on the order of $\mu$~K.
In the next section, we evaluate the anisotropic signal in more detail.

\section{THERMAL SUNYAEV-ZEL'DOVICH ANISOTROPY DUE TO PBHs}

As shown in the previous section,
a hot plasma is generated around a PBH
and can cause the SZ effect.
If PBHs account for a significant fraction of the dark matter
abundance, they can produce the observable CMB temperature anisotropy
through the SZ effect.
In this section, we evaluate the angular power spectrum of the CMB temperature fluctuations by the SZ effect. We also discuss the constraint on the PBH abundance from the comparison with the
small-scale CMB measurement from the SPT.

\subsection{The CMB temperature angular power spectrum by the SZ effect}

To calculate the angular power spectrum of the CMB temperature
due to the SZ effect caused by PBHs, we take a
similar method to the one for galaxy clusters based on the halo
formalism~\cite{Komatsu&Kitayama_SZ,2000MNRAS.318..203S}.
Accordingly,
the angular power spectrum can be described as the sum of the two contributions:
\begin{equation}
C_{l}^{TT}=g^2(\nu) \left(C_{l}^{yy(1 {\rm P})}+C_{l}^{yy(2{\rm P})}\right),
\label{cleq} 
\end{equation}
where 
$C_{l}^{yy(1{\rm P})}$ is the ``one-PBH'' term describing the Poisson contribution
and $C_{l}^{yy(2{\rm P})}$ is the ``two-PBH'' term arising from clustering of PBHs.
Assuming that
the PBH mass function is restricted to a single mass~$M$,
we can write these two terms as 
\begin{align}
C_{l}^{yy(1{\rm P})}=&\int^{z_{\rm{ini}}}_{z_{\rm{f}}}
dz\frac{d^2V}{dzd\Omega}n_{\rm{PBH}} \left|
y_{l}(z)\right|^{2},
\label{1pbh}\\  
C_{l}^{yy(2{\rm P})}=&\int^{z_{\rm{ini}}}_{z_{\rm{f}}}
 dz\frac{d^2V}{dzd\Omega}P\left(\frac{l}{d(z)} \right)
 n_{\rm{PBH}}^2
 \left| y_{l}(z)\right|^{2},
\label{2pbh} 
\end{align}
where $V$ is the comoving volume, $d(z)$ is the comoving distance
to the redshift~$z$,
$P(k)$ is the matter power spectrum and $n_{\rm{PBH}}$ is the comoving number
density of PBHs with mass~$M$.
Here we assume that
PBHs contribute to the DM abundance
with the fraction~$f_{\rm PBH}$.
Therefore, $n_{\rm{PBH}}$ can be written as
$n_{\rm{PBH}}=f_{\rm{PBH}}\Omega_{\rm{DM}} \rho_{\rm crit}/M$
with the present critical density
of the Universe,~$\rho_{\rm crit}$.
We integrate these equations from the redshift $z_\mathrm{ini}$ to $z_\mathrm{f}$ for whose values we will add small discussion later.

In Eqs.~\eqref{1pbh}~and~\eqref{2pbh},
taking the small angle approximation,
we get $y_{l}$ as the two-dimension Fourier transform of the Compton $y$-parameter
for PBH mass $M$ at the redshift~$z$ obtained in Sec.~\ref{sz_by_apbh}:
\begin{equation}
y_{l }(z)=\int d^{2}\theta~y(b){\exp}(-i\bm{\theta}\cdot\bm{l}),
\label{eq:yl-def}
\end{equation}
where $\bm l$ is a vector describing the two-dimensional Fourier mode with $l=|{\bm l}|$, the comoving distance $b$ on the sky
is given in $b = |{\bm \theta}|d(z)$ and
$\bm \theta$ is the angular direction on the sky sphere.

\begin{figure}[tbp]
\centering
\includegraphics[width=8cm,clip]{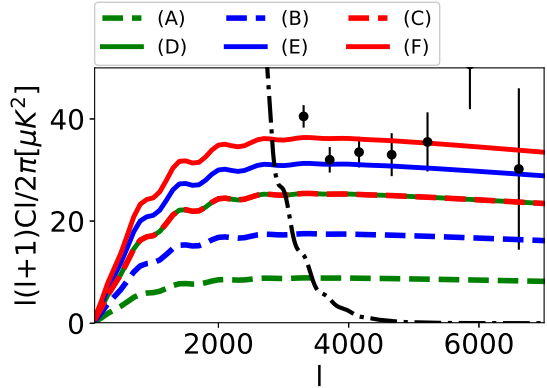}
\caption{The angular power spectrum of the CMB temperature caused by the SZ
 effects from PBHs. The parameter sets for cases (A)ー(F) are as shown in Table~\ref{6case}. The black
 dashed-dotted line is the primordial power spectrum of the CMB temperature
 and the black circles with error bars represent the SPT data.
}
\label{cl} 
\end{figure}

Calculating Eqs.~\eqref{1pbh}~and~\eqref{2pbh} with the profile of
$y(b)$,
we can obtain the CMB temperature anisotropy induced
by the SZ effect due to PBHs.
In Fig.~\ref{cl}, we plot the obtained angular power spectra of the CMB
temperature anisotropy for the different PBH parameter
sets~$(M,~\epsilon,~f_{\rm PBH})$.
Our PBH parameter sets are summarized in Table~\ref{6case}.
For comparison, we show the primordial CMB temperature anisotropy
as the black dashed-dotted line.
We also plot the SPT data with the error bars~\cite{SPT}.

The shape of the spectrum is independent on the PBH parameters.
We find out that the contribution from the two-PBH term
dominates the one-PBH term. 
Since the typical scale of the $y$-parameter profile in Fig.~\ref{y} is
the Str\"omgren radius,~$R_{\rm s}$, which is much smaller
than the CMB observation scales, $y_l$ is constant
on the CMB observation scales. Therefore, according to Eq.~\eqref{2pbh}, 
the shape of the angular spectrum is determined by the matter power spectrum.
In fact, independently on the PBH parameters,
Fig.~\ref{cl} shows that 
the spectrum has a flat shape on larger multipoles than $l \sim 2000$.
Because of this flat shape, the SZ temperature anisotropy due to PBHs can dominate the primordial temperature anisotropy on smaller
scales than the Silk scale.

On the other hand,
the amplitude depends on the PBH parameters.
Because of the integration range in Eq.~\eqref{yparam},
$y(b)$ is proportional to $R_{\rm s}$.
Figure~\ref{y} tells us that 
the typical scale of non-negligible $y(b)$ is also $R_{\rm s}$.
Therefore, according to Eq.~\eqref{eq:yl-def},
the angular Fourier component of the $y$-parameter, $y_l$,
is proportional to $R^{3}_{\rm s} \propto \epsilon M$.
The PBH number density is $n_{\rm PBH} \propto f_{\rm PBH}/M$.
As a result, 
$C_{l}$ has the dependence
as
$
C_{l} \simeq C_{l}^{yy(2{\rm P})}
\propto |n_{\rm{PBH}}y_{l}|^2\propto\ \left(
\epsilon f_{\rm{PBH}} \right)^2
$
and
we found that the angular power spectrum on the flat region~$l>2000$ can be approximated in
\begin{equation}
l^2C_{l} 
\sim 0.5\times
\left(\frac{f_{\rm{PBH}}}{10^{-2}}\right)^2
\left(\frac{\epsilon}{10^{-4}}\right)^2.\label{amp_cl}
\end{equation}

Figure~\ref{cl} clearly shows this dependence of $C_{l}$ to us. 
Because of the degeneracy between $\epsilon$ and $f_{\rm PBH}$,
the angular spectrum for the case C is coincident with that for the case D.

We also investigate the redshift contribution to the angular spectrum.
The observed temperature anisotropy is a resultant effect which is
integrated along the line-of-sight direction.
As already mentioned, $y_{l}$ is proportional to $R_{s}^{3}$ which becomes
large as the Universe evolves.
In particular, we find out that the contribution from the redshift
higher than $z=50$ is less than 1 \%.
On the other hand, in the low-redshift side,
the Universe is gradually reionized and finally
fully ionized until $z\sim 7$~\cite{Planck2018_cospara}.
During the epoch of reionization, the regions ionized by PBHs are caught up in the
reionization process by stars and galaxies.
As a result, in lower redshifts around the epoch of reionization,
the SZ effect due to PBHs becomes fainter as the reionization proceeds.
Here, for simplicity, we take into account the PBH SZ effect between
$z_{\rm ini} =200$ and
$z_{\rm f}=10$ in the integration of Eqs.~\eqref{1pbh} and~\eqref{2pbh}.

In Ref.~\cite{Kamionkowski&Alihaimoud_pbhlimit}, they calculated the typical value of $\epsilon$ in the analytic model with the spherical accretion on a PBH.
Their obtained $\epsilon$ is on the order of $10^{-9}$
for the PBH mass $100M_{\odot}$ at $z\sim 10$.
Substituting this value into Eq.~\eqref{amp_cl}, we can see that the signal is much smaller than the current observation sensitivity
such as the SPT even for $f_{\rm PBH}=1$. 
However, there exists
the theoretical uncertainty in 
$\epsilon$ as mentioned in Sec.~\ref{section2}. 
Depending on the environment around a PBH, 
$\epsilon$ could become larger than the above one.
In that case, we might get the observable value in the present observation status.
Therefore, in the next subsection, 
we discuss the
possible constraint on the PBH abundance 
from the thermal SZ angular power spectrum
with the free parameter $\epsilon$.

\begin{table}[tb]
\begin{center}
  \begin{tabular}{|c|c|c|c|} \hline
 &   $M[M_{\odot}]$ & $\epsilon$ & $f_{\rm PBH}$ \\ \hline \hline
 (A) &  10 & $1.0\times10^{-4}$ & $7.6\times10^{-2}$ \\ \cline{1-4}
 (B)&   100 & $1.0\times10^{-4}$ & $7.6\times 10^{-2}$  \\ \cline{1-4}
 (C)&   1000 & $1.0\times10^{-4}$& $7.6\times10^{-2}$  \\ \hline
 (D)&   10& $1.0\times10^{-2}$ & $7.6\times10^{-4}$  \\ \cline{1-4}
 (E)&   100 &$1.0\times10^{-2}$ & $7.6\times10^{-4}$ \\ \cline{1-4}
 (F)&   1000 & $1.0\times10^{-2}$ & $7.6\times10^{-4}$ \\ \hline
  \end{tabular}
  \caption{The PBH parameter sets for the CMB anisotropy due to PBHs
in Fig.~\ref{cl}.}  
  \label{6case}
        \end{center}
\end{table}

\subsection{Application to the PBH constraint}

Now we discuss the possible constraint on the PBH abundance from the CMB SZ power
spectrum measurement.
When we fix the PBH mass $M$ and the emission efficiency $\epsilon$,
the amplitude of the SZ spectrum is determined by the PBH abundance~$f_{\rm PBH}$ in
our model.

Since the SZ effect due to PBHs can induce the CMB temperature
anisotropy on small scales, the small-scale CMB measurement
provides the constraint on the PBH abundance.
The SPT data reach the minimum value, $C_l^{\rm SPT}$ 
at $l_{\rm SPT} =3709$.
Therefore, the condition, $C_l^{TT} < C_l^{\rm SPT}$ at $l=l_{\rm SPT}$
gives the limit on $f_{\rm PBH}$ with
fixing $\epsilon$ and $M$, because of the flat shape of the spectrum.
We found that the constraint from the SZ angular power spectrum
is roughly independent on the PBH mass and
given in
\begin{equation}
 f_{\rm PBH} < 10^{-3} \left(\frac{\epsilon}{10^{-2}}\right)^{-1},\label{fpbh_ineq}
\end{equation}
for $10^{-2} M_\odot < M < 10^3 M_\odot$.
Therefore, when we adopt $\epsilon$, $10^{-13}<\epsilon < 10^{-7}$ for $1M_\odot<M< 10^3M_\odot$, as in Refs.~\cite{ROM,Kamionkowski&Alihaimoud_pbhlimit},
the current SPT CMB measurement on small scales Eq.~\eqref{fpbh_ineq} 
cannot provide the constraint on the PBH abundance.
When the PBH has $\epsilon $ larger than
$10^{-5}$ as discussed in Refs.~\cite{Poulin:2017bwe,2018JCAP...05..017B},
we can obtain the abundance constraint from the SPT data.
However, 
according to Refs.~\cite{ROM,Kamionkowski&Alihaimoud_pbhlimit}, we found that the CMB measurement of the global Thomson optical depth provides the PBH abundance constraint $f_{\rm PBH}<10^{-9}(\epsilon/10^{-2})^{-1}$ in terms of $\epsilon$.
Therefore, we conclude that the constraint from the SZ measurement is always 
weaker than the one from the global Thomson optical depth.

\section{CONCLUSION}

In this paper, we have investigated the CMB temperature angular power
spectrum due to the SZ effect caused by PBHs.
The gas accreting on a PBH heats up because of the release of the
gravitational energy. As a result, the heated gas in the vicinity of the PBH
emits the UV and x-ray photons.
These photons can ionize and heat the IGM around the PBH.
The ionized hot IGM causes the secondary CMB temperature anisotropy
through the SZ effect.

First assuming the luminosity of photons
emitted in the vicinity of a PBH with a free parameter,
we have evaluated the profiles of the IGM ionization fraction
and temperature around a PBH by solving the 
one-dimensional radiative transfer equations.
Based on these profiles, we have obtained the SZ Compton
$y$-parameter profile around a PBH.
Following the halo formalism, finally, we have calculated the CMB temperature
angular power spectrum due to the PBH SZ effect with assuming the PBH abundance.

We have shown that the SZ spectrum due to PBHs
could dominate the primordial temperature spectrum on
smaller scales than the Silk scale. 
Since the correlation on such scales is made by 
$y$ distortions between two different PBHs,
the shape of the angular power spectrum depends on 
the matter power spectrum.
We have also found that the amplitude of the spectrum
is sensitive to the PBH abundance and the emission efficiency of
the gas accreting on PBHs.
In the case of a scale-invariant primordial spectrum,
the angular power spectrum $l^2 C_l$ has a flat shape on small scales
$l^2C_{l} 
\sim 0.5\times\left(f_{\rm{PBH}}/10^{-2}\right)^2\left(\epsilon/10^{-4}\right)^2$.
Using $\epsilon$ in Ref.~\cite{Kamionkowski&Alihaimoud_pbhlimit}, we cannot get the observable signal in current observations such as the SPT.
However, $\epsilon$ has a theoretical uncertainty and Refs.\cite{Poulin:2017bwe,2018JCAP...05..017B} suggest that PBHs could have $\epsilon$ larger than $10^{-5}$.
In this case, PBHs could make the observable signature.

We also investigated the impact for the constraint of the PBH abundance from the SZ effect.
Our obtained constraint on the PBH fraction to dark matter 
is $f_{\rm PBH} \lesssim 10^{-3} (\epsilon / 10^{-2})$
for the PBH mass range $10^{-2} M_{\odot} <M<10^3 M_{\odot}$.
However, we should note that the CMB constraint from the optical depth provides a much tighter constraint
with the same $\epsilon$.
Therefore we conclude that the SZ measurement does not give a new constraint on the PBH abundance.

Although the constraint from the SZ angular power spectrum is weaker than that from the optical depth,
it is worth mentioning the impact of the future small-scale SZ
measurements on the probe of the existence of PBHs.
We have shown that the CMB temperature spectrum of the SZ effect due to
PBHs has the flat spectrum on smaller scales than $l \sim 2000$.
This flat spectrum extends up to the scale of the ionized region by the PBH
emission, e.g., roughly 1~kpc in the physical scale for $M=10M_{\odot}$ with $\epsilon
=10^{-4}$.
The SZ effect in galaxy clusters can produce
large CMB anisotropies on small scales.
However, its spectrum has a peak
around $l \sim 4000,$ and the amplitude decays on higher $l$.
Therefore, the detection or nondetection of the flat SZ spectrum on higher $l$
than $l=4000$ gives
useful information about the existence of PBHs.

In this work, we have assumed the luminosity from the hot gas in the
vicinity of a PBH with introducing a free parameter for the emission
efficiency.
References.~\cite{ROM,Kamionkowski&Alihaimoud_pbhlimit,Poulin:2017bwe} have studied
the emission efficiency 
with the simple assumptions analytically.
However,
to obtain the emission efficiency consistently,
we need to solve the dynamics of gas accreting on a PBH
with considering the backreaction of the luminosity from hot gas.
In the next study,
we address this issue by numerical simulations
and investigate the cosmological effects of the emission from gas
accreting on a PBH in more detail. 

\begin{acknowledgments}
We are grateful to Kenji Hasegawa for his useful comments.
This work is supported by JSPS KAKENHI Grants No. 15K17646 (H.T.) and
No. 17H01110 (H.T.).
\end{acknowledgments}

\bibliography{article3}

\end{document}